\documentclass[12pt]{article}

\usepackage{graphicx}
\usepackage{psfrag}   
\newcommand{\be}[0]{\begin{equation}}
\newcommand{\ee}[0]{\end{equation}}
\newcommand{\bea}[0]{\begin{eqnarray}}
\newcommand{\eea}[0]{\end{eqnarray}}
\newcommand{\nn}[0]{\nonumber}

\begin{document}
\large
\hfill\vbox{\hbox{DCPT/06/54}
            \hbox{IPPP/06/27}}
\nopagebreak

\vspace{2.0cm}
\begin{center}
\LARGE
{\bf { Crouching Sigma, Hidden Scalar\\ [3mm]
in $\gamma\gamma\to\pi^0\pi^0$ }}

\vspace{12mm}

\large{M.R. Pennington}

\vspace{5mm}
{\it Institute for Particle Physics Phenomenology,\\ Durham University, 
Durham, DH1 3LE, U.K.}
\end{center}
\vspace{1cm}

\centerline{\bf Abstract}

\vspace{3mm}
\small

There has long  been speculation about the nature of the $\sigma$-resonance. For three decades Jaffe has argued for a tetraquark
composition, while others have claimed it is largely a glueball. 
A key pointer to its nature is its coupling to two photons. Consequently, there
have been recent proposals to observe this important scalar hiding in $\gamma\gamma\to\pi^0\pi^0$. We show here that
the $\sigma$ is already crouching in this cross-section exactly as measured twenty years ago. 
What is new is that precise knowledge of the position of the $\sigma$-pole, provided by  the 
analysis of the Roy equations, now allows its two photon coupling to be accurately fixed. 
Its two photon width is found to be $(4.1 \pm 0.3)$ keV, a value far too large for a gluonic, or even a tetraquark, state.

\normalsize

\newpage

\parskip=2.5mm
\baselineskip=6.5mm

\section{New information on the $\sigma$}

The $\sigma$-resonance has for long been a mnemonic for the highly correlated two pion exchange that generates the longest range isoscalar force revealed in  nuclear binding. It is also the name of the scalar field, the non-zero vacuum expectation value of which, breaks chiral symmetry: giving mass to all light hadrons
~\cite{mrp-mexico}. While isoscalar $\pi\pi$ interactions grow rapidly above threshold, they have none of the features readily identified as a text-book resonance, quite unlike the $\rho$ for instance. If $\pi\pi$ mass distributions, whether from classic meson-meson reactions or from final state interactions of decay products, are fitted with Breit-Wigner forms, then inevitably one finds a pole in the complex energy plane.
However, fits give a position varying wildly from one analysis to another with both masses and widths from 350 MeV to 1 GeV~\cite{PDG,MRPwhs}.

\noindent  Renewed interest in using  the Roy equations, which encode the analyticity provable in axiomatic field theory with the three channel crossing symmetry of $\pi\pi$ scattering,  has, when combined with chiral constraints and new experimental information, allowed a narrow corridor of possible amplitudes from 800 MeV down to threshold, as found by Colangelo, Gasser and Leutwyler~\cite{cgl}.
Recent recognition that the Roy equations can be evaluated not just on the real axis but in the complex energy plane has determined the position of the lightest resonance in QCD, the  $\sigma$, to be at $E_R\,=\,441\,-\,i\,272$ MeV within small uncertainties~\cite{caprini}. 
But what is the nature of this state in the spectrum of hadrons? Is it a conventional ${\overline q}q$ state of the quark model~\cite{vanbeveren}? Is it a tetraquark meson~\cite{jaffe}, composed of $\overline{qq}qq$, with the expected ${\overline q}q$ nonet still higher in mass~\cite{schechter,maiani}, or is it largely glue~\cite{narison,minkochs}?
The coupling to photons is a key guide to a state's composition.

\noindent  Now the same precise information on $\pi\pi$ amplitudes that determines the existence and the position of the pole allows the amplitude for $\gamma\gamma\to\pi\pi$ to be accurately determined. Exploiting this is most readily done by the use of partial wave dispersion relations. For a hadronic scattering process the analyticity 
needed to deduce such relations is not provable in axiomatic field theory being limited by our understanding of the Mandelstam double spectral region. However, for the electromagnetic process $\gamma\gamma\to\pi\pi$, these are on firmer footing and this is what we require.
While the $\sigma$ can only appear in the $I=0$ channel, we need to consider the $I=2$ amplitude at the same time. This is because both isospins contribute 
with almost equal importance to both the $\pi^+\pi^-$ and $\pi^0\pi^0$ channels, as we will emphasise later.

\section{Two photon amplitude}

  Let us begin by considering the $S$-wave $\gamma\gamma\to\pi\pi$  amplitudes with isospin $I$, ${\cal F}^I(s)$, where $s$ is the square of the $\pi\pi$ invariant mass. Each of these amplitudes, with $I=0,\,2$, being complex have a phase $\phi^I(s)$ along the right hand cut, when $s$ is above the two pion threshold, {\it i.e.} $s\, >\, s_{th}\,=\, 4m_{\pi}^2$.
Unitarity, through Watson's theorem, requires the phase of each of these partial waves to be
 the same~\cite{pi} 
as the phase  of the corresponding $\pi\pi$ partial wave amplitude with the same spin and isospin in the elastic region. 
 To implement this constraint we define the Omn\`es function, $\Omega^I(s)$, by
\be
\Omega^I(s)\;=\;\exp \left[ \frac{s}{\pi}\,\int_{s_{th}}^{\infty}\;
ds'\,\frac{\phi^I(s')}{s'(s'-s)} \right]\, \quad .
\ee
which by construction has phase $\phi^I(s)$ for $s > s_{th}$.
Thus the $\gamma\gamma\to\pi\pi$ $S$-wave amplitudes, ${\cal F}^I(s)$,
can be written as $P^I(s)\,\Omega^I(s)$, where $P^I(s)$ is a function which is real along the right hand cut with $s > s_{th}$. 
The phase, $\phi^I$, is simply the phase-shift in the region of elastic unitarity, which applies up to ${\overline K}K$ threshold, since multi-pion channels are negligible below 1.2 GeV. 
Moreover, in the low energy region of interest where $|\,s\,|\,\sim\, 0.25$ GeV$^2$, the differences in phase above 1 GeV affect the results little as has been checked by replacing the $\pi\pi\to\pi\pi$ phase with that for $\pi\pi\to {\overline K}K$. Such a change is equivalent to assuming the $\pi\pi$ final state in the two photon process is only accessed through a ${\overline K}K$ intermediate state. Outside the narrow confines of the $f_0(980)$ region, this would be an extreme possibility. Nevertheless, the effect is small and included in the uncertainties we quote.
Representative input $\pi\pi$ $S$-wave phases, $\phi^I(s)$, for $I=0,\,2$ and the resulting Omn\`es functions are shown in Fig.~1.
\begin{figure}[t] 
\begin{center}
\includegraphics[width=12.cm]{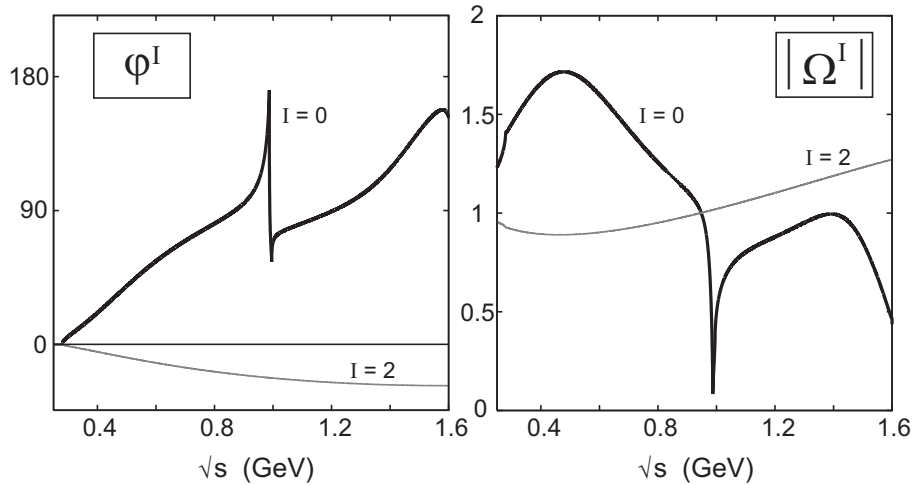}
\end{center}
\caption{\leftskip=1.cm\rightskip=1.cm{Representative $I=0,\,2$ $\gamma\gamma\to\pi\pi$ $S$-wave phases and moduli of the  Omn\`es functions, $\Omega ^I(s)$, related by  Eq.~(1).}}
\end{figure}

\noindent  Now Low's low energy theorem~\cite{low} requires that as $s \to 0$, and $t,u \to m_{\pi}^2$, at the threshold for Compton scattering $\gamma\pi\to\gamma\pi$, the full scattering amplitude is equal to its one pion exchange Born term.
It is such crossed channel exchanges that generate the  left hand cut contribution to the 
$\gamma\gamma\to\pi\pi$ partial wave amplitudes, which we denote collectively
by ${\cal L}^I(s)$. Because the pion is so much lighter than any other hadron,
 pion exchange determines the discontinuity across this left hand cut not just at $s=0$ but in the whole region $0\,>\,s\,>\,-M_V^2$, beyond which other exchanges like $\rho, \omega$ start to contribute~\cite{mv}. While the Born term assumes pointlike couplings for the pion, any form-factor dependence only affects the left hand cut for $s\,<\, -M_V^2$, since it is vector masses that set the scale for such charged radii. Consequently, the left hand cut from $s=0$ to $s\,\simeq\, -0.5$ GeV$^2$ is precisely known and that is all we require to fix the amplitude in the region of $s\,=\,s_R\,=\,E_R^2\,$ shown in Fig.~2. 
\begin{figure}[h] 
\begin{center}
\includegraphics[width=12.cm]{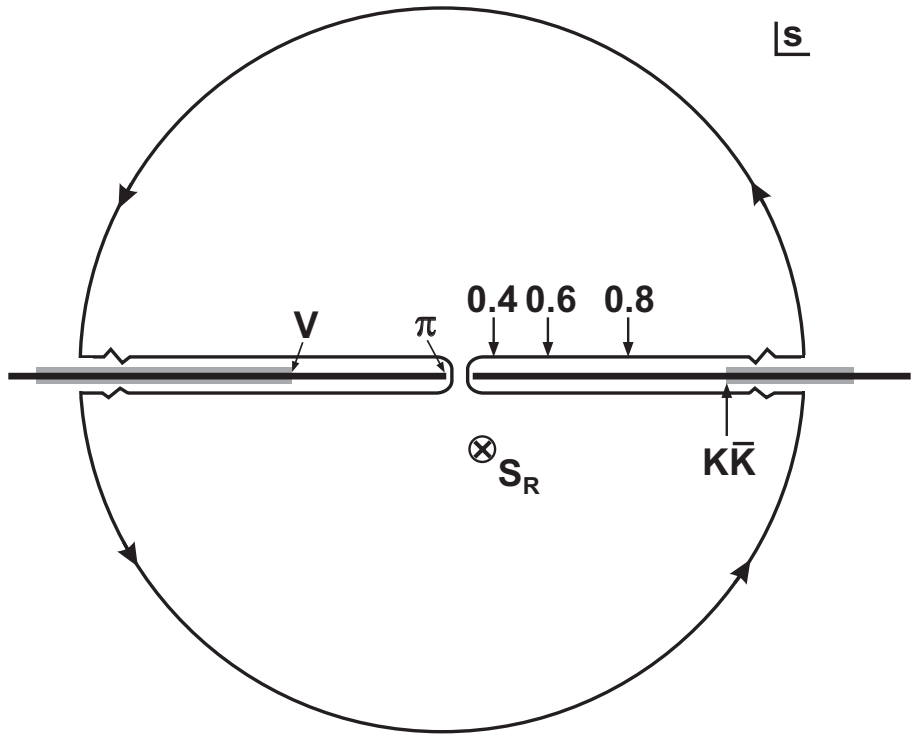}
\end{center}
\caption{\leftskip=1.cm\rightskip=1.cm{The complex $s$-plane structure of the $\gamma\gamma\to\pi\pi$ amplitudes, ${\cal F}^I(s)$. $\pi$ labels the start of the left hand cut controlled by the pion exchange Born term, while $V$ denotes where the vector exchanges $\rho, \omega$ start to contribute to the discontinuity. The right hand cut is elastic effectively up to ${\overline K}K$ threshold. The point $s=s_R$ is the position of the $\sigma$ pole~\cite{caprini}. The plot is drawn to scale so 0.4, 0.6, 0.8 are the c.m. energy in GeV.}}
\end{figure}

\newpage
 
\noindent  To see how, let us construct
 the function $G^I(s)\;\equiv\;\left({\cal F}^I(s)\,-\,{\cal L}^I(s)\right)\,\Omega^I(s)^{-1}$, which only has a right hand cut. Its
discontinuity is ${\cal L}^I(s)\,\sin \phi^I(s)/| \Omega^I(s) |$, which is accurately known at low energies.  This information is embodied in a dispersion relation for the
function $G^I(s)$ using a contour like that in Fig.~2.
 While the behaviour of $G^I(s)$ means the  integral at infinity converges with just one subtraction, it is more convenient  for our purpose to ensure 
that the integrals are dominated
by the known low energy regime of $|\, s'\, |\, <\, M_{\rho}^2$. This is achieved by making two subtractions:
\bea
\nn
{\cal F}^I(s)\;=&&{\cal L}^I(s)\,+\,c_I s\, \Omega^I(s)\, \\
&&+\,\frac{s^2}{\pi}\,\Omega^I(s)\,\int_{s_{th}}^{\infty}\, ds'\,\frac{{\cal L}^I(s')\,\sin \phi^I(s')}{s'^2\,(s' - s)\,|\,\Omega^I(s')\,|}\quad ,
\eea
The subtraction constants $c_I$ are specified by the QED low energy theorem and chiral dynamics.
These two conditions apply to the amplitudes with the pions of definite charge (which are combinations of those with definite isospin).
Low's theorem, as shown by Goldberger {\it et al.}~\cite{goldberger} requires that the $S$-wave amplitude for $\gamma\gamma\to\pi^+\pi^-$, 
\be
{\cal F}^{+-}(s)\;\to\; B(s)\, +\,{\cal O}(s^2)\quad {\rm as}\quad  s \to 0 ,
\ee 
where $B(s)$ is the Born $S$-wave, while chiral dynamics requires that the $S$-wave amplitude for $\gamma\gamma\to\pi^0\pi^0$
\begin{figure}[t] 
\begin{center}
\includegraphics[width=10.cm]{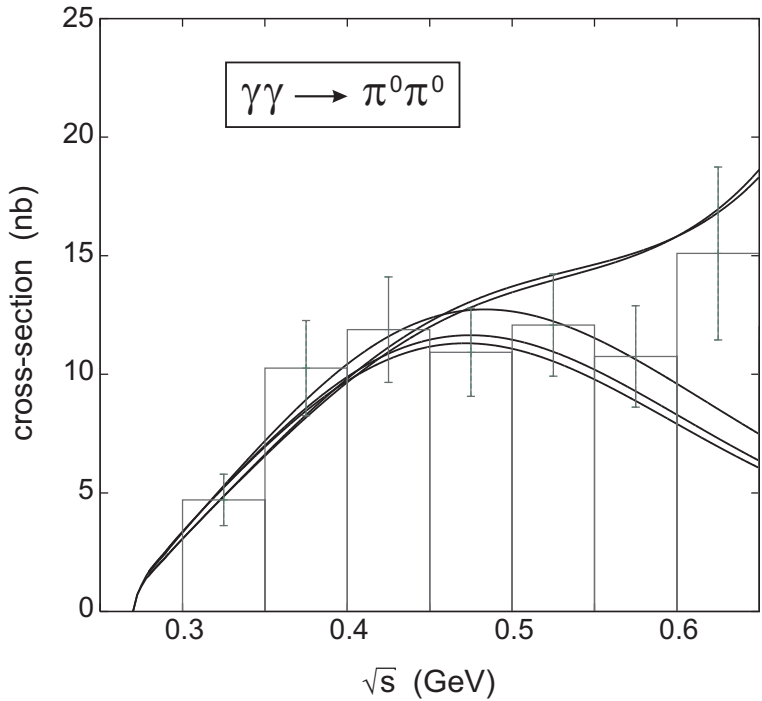}
\end{center}
\caption{\leftskip=1.cm\rightskip=1.cm{Results of the dispersive calculation for the low energy $\gamma\gamma\to\pi^0\pi^0$ cross-section for different input phases $\phi^I$ above ${\overline K}K$ threshold, each with 3 different positions of the Adler zero, Eq.~(4), at $s = 1/2, 1, 2 m_{\pi}^2$, compared with the Crystal Ball data~\cite{CB} scaled to the whole angular range.}}
\end{figure}
\be
{\cal F}^{00}(s)\;=\;0\quad
 {\rm at}\quad s\,=\,{\cal O}(m_{\pi}^2)\quad .
\label{sn}
\ee
At one loop level in Chiral Perturbation Theory~\cite{bijnens,donoghue},
$
{\cal F}^{00}(s)\;\propto \,{\cal T}(\pi^+\pi^-\to\pi^0\pi^0)
$
and so places the Adler zero exactly at $s\,=\,m_{\pi}^2$ at this order. However, its precise position hardly affects our results. 

\noindent
  These relations allow us to determine the $\gamma\gamma\to\pi\pi$ cross-section in the low energy region. Precision comes from the more accurate determination of the $\pi\pi$ $S$-wave amplitudes obtained by combining new results from decays like $K_{e4}$, $J/\psi \to \phi X$ and $D\to\pi X$~\cite{boglione} with the Roy equations. 
This calculation reproduces the cross-section for the production of charged and neutral pions as measured by Mark II~\cite{MarkII} and Crystal Ball~\cite{CB}, respectively, in the low energy region with no free parameters. The predictions for the neutral cross-section are shown in Fig.~3. The range shown delineates the uncertainties due to (i)~different $\gamma\gamma$ phases $\phi^I(s)$ above ${\overline K}K$ threshold and (ii)~different positions of the Adler zero in Eq.~(4).  Notice that the cross-section is very  nearly  unique up to 450 MeV.

\noindent
Of course, the $I=0$ $\pi\pi$ phase and Omn\`es function, shown in Fig.~1, 
know about the $\sigma$-pole at $s=s_R$
deep in the complex plane close to both the right and left hand cuts of Fig.~2.

\section {Two photon coupling of the $\sigma$}

  Not only can we determine the $\gamma\gamma$ amplitudes ${\cal F}^I(s)$ along the upper side of the right hand cut on the physical sheet where experiments are performed, but everywhere on this first sheet. In particular, we can determine the $I=0$ amplitude 
at $s\,=\,s_R$, marked in Fig.~2.
The right hand cut structure of the $\gamma\gamma\to\pi\pi$ amplitude mirrors that of the corresponding hadronic amplitude, ${\cal T}^I$, for  $\pi\pi\to\pi\pi$ in the region of elastic unitarity:
\be
{\cal F}^I(s)\;=\;\alpha^I(s)\;{\cal T}^I(s)\quad ,
\ee
where the function $\alpha^I(s)$ represents
the intrinsic coupling of $\gamma\gamma\to\pi\pi$, while ${\cal T}$ describes the final state interactions, which colour and shape the electromagnetic process
~\cite{amp}.

\noindent   At $s\,=\,s_R\,$ on the first sheet, the amplitude ${\cal T}^{I=0}(s)\,=\,i/2\rho(s)$, since the $S$-matrix element vanishes at this point. $\rho(s)$ is, as usual, the phase-space factor
$\rho(s)\,=\,\sqrt{1 - s_{th}/s}$. The dispersion relation on the first sheet then determines
the coupling function $\alpha(s_R)$, which not having a right hand cut, has the same value on the second sheet. We introduce subscripts to label the sheets $I$ and $II$, while the superscripts continue to denote isospin.

\baselineskip=6.2mm

\noindent  In the neighbourhood of the pole on the second sheet, the $\gamma\gamma\to\pi\pi$ $S$-wave amplitude is given by
\be
{\cal F}^0_{II}(s) \;\simeq\;\frac{g_{\gamma}\,g_{\pi}}{s_R - s}\; ,\quad{\rm while}\quad
{\cal T}^0_{II}(s) \;\simeq\;\frac{g_{\pi}^2}{s_R - s}\; .
\ee 
Thus $\alpha^0(s_R)$ determines the ratio of $g_{\gamma}/g_{\pi}$ for the isoscalar resonance. 
Now the hadronic amplitude on sheet I is related to that on sheet II by
\be
\frac{1}{{\cal T}_{II}(s)}\;=\;\frac{1}{{\cal T}_{I}(s)}\,+\,2\,i \rho\quad ,
\ee
so that
\be
g_{\gamma}^2\;=\;\lim_{s\to s_R}\;\frac{(s - s_R)\,F^0_{I}(s)^2}{\left({\cal T}^0_{I}(s)\,-\,i/2\rho\right)}\quad .
\ee
Combining the representation cited above~\cite{boglione} for the hadronic amplitude, ${\cal T}^0$, on sheet I with the present dispersive calculation then gives
the two photon coupling of the $\sigma$ , which specifies its radiative width to be~\cite{morpenngg}
\be
\Gamma(\sigma\to\gamma\gamma)\;=\; \frac{\alpha^2 |\,\rho(s_R)\,g_{\gamma}^2\,|}{4 M_{\sigma}}\,=\,(4.09\,\pm\,0.29)\; {\rm keV}\;.
\ee

\noindent  That this is ten times larger than the signal seen in $\gamma\gamma\to\pi^0\pi^0$ cross-section requires some explanation, particularly in the light of proposals, {\it e.g.}~\cite{nguyen}, to search for the $\sigma$ in this channel. If 
we consider this process and for the moment completely ignore the requirement that final state interactions shape the $\pi\pi$ distribution in a well-defined way. Then one would say there is no Born contribution to this channel
and so the cross-section should reflect the appearance of resonant structures if they exist. If this is the $\sigma$, then one can read off from the observed cross-section in Fig.~3 of 10-12 nb  a $\gamma\gamma$ width an order of magnitude smaller than we have deduced.

\noindent In hadronic channels $I=2$ amplitudes, which are {\it exotic} in the quark model, are much smaller than those with $I=0$. In contrast in this two photon
process both $I=0,\,2$ are equally important. The $\sigma$  appears in the $I=0$ amplitude, and this can only be separated from data by analysing $\gamma\gamma\to\pi^+\pi^-$ and $\pi^0\pi^0$ together~\cite{morpenngg}.

\begin{figure}[h] 
\begin{center}
\includegraphics[width=10.cm]{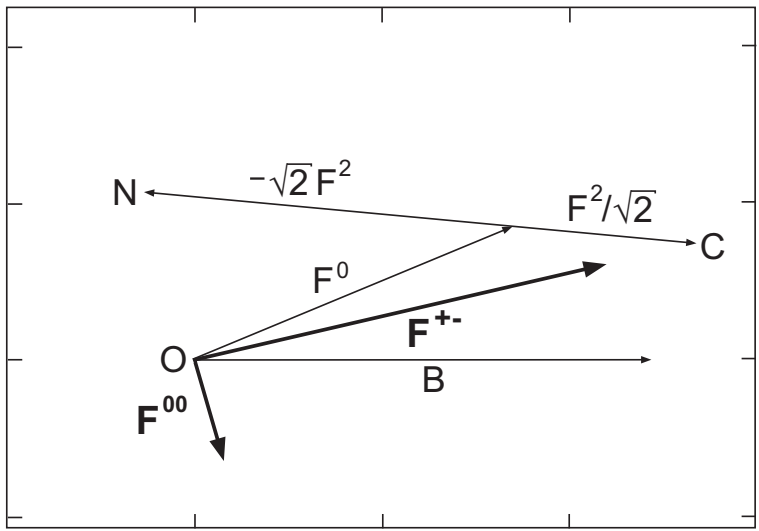}
\end{center}
\caption{\leftskip=1.cm\rightskip=1.cm{$\gamma\gamma\to\pi\pi$ $S$-wave amplitudes at 400 MeV with definite isospin and with definite charges as indicated by the superscripts. $B$ is the Born amplitude $S$-wave for comparison. $OC$ and $ON$ define the directions of the charged and neutral pion amplitudes as given by the vector sums described in the text.}}
\end{figure}
 
\noindent
As we have seen what is really happening is that the Born amplitude is modified by final state interactions to ensure Watson's theorem is satisfied. As a result the
$I=0$ and $I=2$ amplitudes are no longer real and exactly cancelling in the neutral channel. A vectorial representation of this is shown at 400 MeV in Fig.~4.
The $I=0$ component has the phase of $I=0$ $S$-wave $\pi\pi$ scattering, while that with $I=2$ has the phase of the corresponding isotensor $S$-wave.
In Fig.~4 the vector $OC$ $({\cal F}^0 + {\cal F}^2/\sqrt{2})$ is $\sqrt{3/2}$ times the charged channel $S$-wave, while the vector $ON$
$({\cal F}^0 - \sqrt{2} {\cal F}^2)$ is $-\sqrt{3}$ times the neutral one. One sees that the square of the neutral channel $S$-wave (which dominates its cross-section) is a factor of 12 smaller than the modulus squared of the $I=0$ $S$-wave. It is in this amplitude with definite isospin that the $\sigma$ is to be found. It is there crouching in the $\pi^0\pi^0$ cross-section. 

\noindent
  A $\sigma$ with a 4 keV coupling to two photons has little of the aspects of a low lying glueball. Rather such a width
is just what Chanowitz~\cite{chanowitz} and Barnes~\cite{barnes} have predicted for an isoscalar state with just
$u$ and $d$ quarks. If it is the tensor companion of the $f_2(1275)$ with a $({\overline u}u + {\overline d}d)/\sqrt{2}$ composition then adapting a positronium result to the non-relativistic quark model, we have the relation:
\be
\Gamma(\sigma\to\gamma\gamma)/\Gamma(f_2\to\gamma\gamma)\;=\;15/4 \times \left(m_{\sigma}/m_{f_2}\right)^n\; ,
\ee
with relativistic corrections estimated in  \cite{rel-corr} to be $\sim 0.5$.
The power $n$ depends on the shape of the potential, being $n=3$ for a Coulomb form. 
With $\Gamma(f_2\to\gamma\gamma)\, \simeq \, 3$ keV~\cite{PDG,belle},  we obtain our calculated radiative $\sigma$ width
with $n \simeq 0.3-1$, perhaps reflecting the long range nature of the binding needed for the $\sigma$.  Such a state, which is very short lived (its total width is $\sim 550$ MeV), inevitably has multiquark components in its Fock space.
However, these may well be more diffuse than any ${\overline q}q$ component and so less able to annihilate readily  into photons. 
In keeping with this, Achasov~\cite{achasov} and Narison~\cite{narison2} predict tetraquark states to have tenths of keV as radiative widths. Thus 4 keV points to a conventional ${\overline u}u$, ${\overline d}d$ composition for the $\sigma$. Hopefully the present result will motivate dynamical calculations to confirm
this is really so.  A 4 keV width is difficult to reconcile with a glueball composition, even in the low energy strong coupling regime. So despite crouching in the cross-section $\gamma\gamma\to\pi^0\pi^0$, the \lq red dragon' of Minkowski and Ochs~\cite{minkochs} is unlikely to be gluonic.    
\vspace{1cm}

\noindent{\bf Acknowledgements}

  I am particularly grateful to Prof. Yasushi Watanabe, not only for the kind hospitality and travel support of Tokyo Institute of Technology, but most importantly for his questioning that initiated this work.
I  acknowledge partial support of the EU-RTN Programme, 
Contract No. HPRN-CT-2002-00311, \lq\lq EURIDICE''.


\begin{thebibliography}{555}
\bibitem{pi} Differences of integer multiples of $\pi$ are eliminated by the 
known threshold behaviour.
\bibitem{amp} This relation generalises beyond the elastic region to ${\cal F}(s)\,=\,\sum_n\,\alpha_n(s)\,{\cal T}(n\to\pi\pi;s)$ where the real coupling functions $\alpha_n$ represent the intrinsic $\gamma\gamma\to$ channel \lq$n$' coupling, and the complete right hand cut is encoded in the hadronic amplitudes for intermediate state $n\to\pi\pi$.
\bibitem{belle} A precision determination of the $f_2$ coupling from new charged pion data from  Belle will appear shortly.
\end{thebibliography}
\end{document}